\documentclass[pre,twocolumn,showpacs,floatfix]{revtex4}
 \usepackage{graphicx}

 \newcommand{\be}{\begin{equation}}
 \newcommand{\ee}{\end{equation}}
 \newcommand{\bea}{\begin{eqnarray}}
 \newcommand{\eea}{\end{eqnarray}}

\begin{document}

 \title{Effect of spatial bias on the nonequilibrium phase transition in a
system of coagulating and fragmenting particles}

 \author{R. Rajesh}
 \email{r.ravindran1@physics.ox.ac.uk}
 \author{Supriya Krishnamurthy}
 \altaffiliation{present address:  Santa Fe Institute, 1399 Hyde Park
Road, Santa Fe, NM 87501,US}
 \affiliation {Department of Physics - Theoretical Physics, University of
Oxford, 1 Keble Road, Oxford OX1 3NP, UK}

 \date{\today}

 \begin{abstract}
 We examine the effect of spatial bias on a nonequilibrium system in which
masses on a lattice evolve through the elementary moves of diffusion,
coagulation and fragmentation. When there is no preferred directionality
in the motion of the masses, the model is known to exhibit a
nonequilibrium phase transition between two different types of steady
states, in all dimensions. We show analytically that introducing a
preferred direction in the motion of the masses inhibits the occurrence of
the phase transition in one dimension, in the thermodynamic limit. A
finite size system, however, continues to show a signature of the original
transition, and we characterize the finite size scaling implications of
this. Our analysis is supported by numerical simulations.  In two
dimensions, bias is shown to be irrelevant.
 \end{abstract}
 \maketitle

 \section{\label{sec1} Introduction}

Systems far from equilibrium can undergo nonequilibrium phase transitions
between two different types of steady states when the parameters of the
system are varied. It is important to know how robust such transitions are
with respect to changes in the governing dynamics and to ask if a
signature of the original phases remains, even if the transition is lost.
The introduction of a spatial bias (a preferred direction)  is one factor
which is known to affect the scaling functions and the exponents
characterizing nonequilibrium transitions.  Examples where bias plays a
role include models of extremal dynamics \cite{bak-sn, mas-zh}, the simple
exclusion process \cite{asep}, sand-pile models \cite{dhar}, directed
percolation \cite{dp}, interface depinning \cite{sneppen},
reaction-diffusion systems \cite{react-diff} and random walkers in fractal
media \cite{biasrw}. These examples consist of systems in which parameters
need to be tuned to reach criticality as well as systems that are
self-organized critical.  In both cases, bias either changes the
universality class characterizing the system
\cite{bak-sn,mas-zh,asep,dhar,dp,sneppen,react-diff} or causes
localization \cite{biasrw} or induces boundary driven transitions
\cite{asep}. 

In this paper we study the effect of bias on a recently introduced model
of aggregation and fragmentation, which was shown \cite{MKB1,MKB2} to
exhibit an unusual nonequilibrium phase transition belonging to a
universality class different from models studied earlier
\cite{mukamel,asep}.  We show that bias in this model plays a different
role to any of the cases mentioned above in that even an arbitrarily small
bias inhibits the phase transition entirely in one dimension. Remarkably
though, a signature of the transition remains and modifies the finite-size
behavior of the system, and we characterize the scaling implications of
this. In two and higher dimensions, we show that bias has no effect on the
phase transition. 

\begin{figure}
 \includegraphics[width=8.0cm]{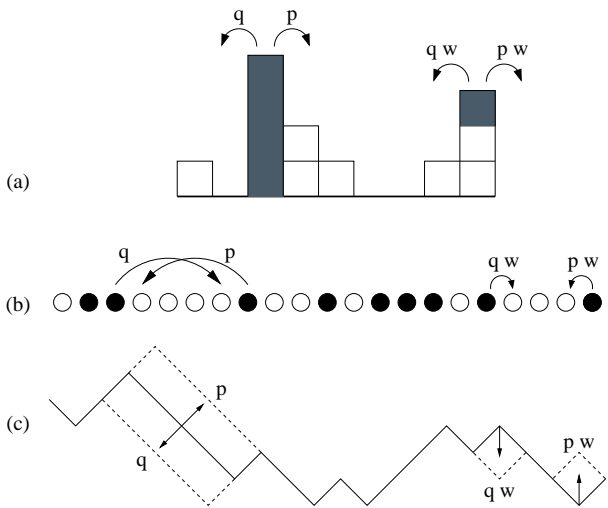}
 \caption{\label{fig0} (a) The model as defined in the text. (b) To obtain
the hard core lattice model from the masses, lay down the masses on their
sides. (c) The interface is obtained by replacing every shaded circle by a
line segment in the $+45^{\circ} $ direction and empty circle by line
segment in the $-45^{\circ} $ direction.}
 \end{figure}

We define the model on a $d$-dimensional hyper-cubic lattice with periodic
boundary conditions.  Starting from a random distribution of non-negative
integer masses, the system evolves in time via the following microscopic
processes. In an infinitesimal time $dt$, (i)  with probability $p dt$ ($q
dt$), the mass at each site hops to one of its nearest neighbors with
increasing (decreasing) coordinates.  (ii) with probability $w p dt$ ($w q
dt$), unit mass is chipped off from an already existing mass and added to
one of its nearest neighbors with increasing (decreasing) coordinates. 
Following these moves, the masses at each site add up (see
Fig.~\ref{fig0}). The dynamics conserves the total mass of the system.
Hence, the parameters defining the system are the bias $p-q$, the density
$\rho$ and the chipping or fragmentation rate $w$.  The case $p=q$
corresponds to the zero bias case or the symmetric model while the case
$p\ne q$ introduces a preferred direction in the motion of the masses and
corresponds to the asymmetric model. 

The model may be mapped onto generalizations of other well studied models
of nonequilibrium statistical mechanics.  In one dimension the system of
masses described above may be equivalently thought of \cite{MKB2} as a
collection of particles and holes on a ring, or a one dimensional
interface evolving in time (see Fig.~\ref{fig0}).  In this language, the
$w\rightarrow \infty$ limit exactly corresponds to the well studied
symmetric (asymmetric) exclusion process \cite{asep_ref} for $p=q$ ($p\neq
q$). In the interface language, this corresponds \cite{interface} to an
interface evolving via the Kardar Parisi Zhang or the Edwards-Wilkinson
dynamics \cite{hal} for the cases $p\ne q$ and $p=q$ respectively.  For a
finite $w$, the nearest neighbor particle exchanges of the exclusion
process (or corner flips of the interface growth model) are further
augmented by long-range moves. 

Models similar to the one studied in this paper have also been studied in
other contexts. A slightly different off-lattice version of this model was
studied within the rate equation approach in the context of polymer chain
growth \cite{KR}. In this case, the mass clusters were thought to
represent polymers. Although the aggregation of polymers and dissociation
of single monomers was allowed in this model, it lacked the important
process of local diffusion which we include in our study. Various models
of coagulation and fragmentation, with coagulation rate proportional to
mass, have been studied in the context of gelation \cite{VZL}.  In these
systems, due to the enhanced coagulation, a gel (a cluster which contains
a finite fraction of the total mass) forms at a finite time. The exponents
at the transition point were shown to depend on whether the process of
fragmentation was present or absent.  Models of fragmentation in which a
fraction of existing mass (as opposed to a single particle) may break off
have also been studied in \cite{KG,RM1,taguchi}. In these models, it could
be deduced that an infinite aggregate that contains a finite fraction of
the total mass never forms. A model with bias very similar in spirit to
the model we study here, was also studied in the context of traffic flow
\cite{IK} and the phase transition observed numerically was interpreted as
a traffic jam occurring for large densities of cars.

An understanding of the steady state reached by the system may be obtained
by considering the limits of only diffusion ($w=0$) or only chipping
($w=\infty$).  Both these cases are exactly solvable. In the former case,
the system maps to the well-studied reaction diffusion system $mA+nA
\rightarrow (m+n)A$, and the steady state reached is simply one in which
all the particles accumulate on a single site.  In the opposite limit ($w
\rightarrow \infty$), the model is again exactly solvable and the system
reaches a steady state in which the mass is uniformly spread out over the
system with the probability that a given site has a mass $m$ being
exponentially distributed.

The special case $p=q$, corresponding to zero bias, was studied earlier
using a mean field approximation \cite{MKB1}, Monte Carlo simulations
\cite{MKB2,RM}, as well as some exact analysis \cite{RM}. It was shown
that the system undergoes a nonequilibrium phase transition in the
$\rho-w$ plane at some critical density $\rho_c(w)$. In particular,
$P(m)$, the probability that a randomly chosen site has mass $m$ was shown
to vary for large $m$ as i) $P(m) \sim \exp (-m/m^*)$ for $\rho<\rho_c$,
ii) $P(m) \sim m^{-\tau}$ at $\rho=\rho_c$ with $\tau=5/2$ and (iii) $P(m)
\sim m^{-\tau} + \mbox{``infinite aggregate''}$ for $\rho> \rho_c$, where
by ``infinite aggregate'' we mean a single large mass equaling a finite
fraction of the total mass of the system.  That is, the mass distribution
$P(m)$ changes from an exponential distribution to an algebraic one at
$\rho_c$. For $\rho > \rho_c$, the mass distribution remains the same as
at $\rho_c$, while all the mass in excess of the critical density
coagulates together forming an infinite aggregate. The mathematical
mechanism giving rise to the infinite aggregate was found to be very
similar to that of equilibrium Bose-Einstein condensation in an ideal Bose
gas. 
 \begin{figure}
 \includegraphics[width=8.0cm]{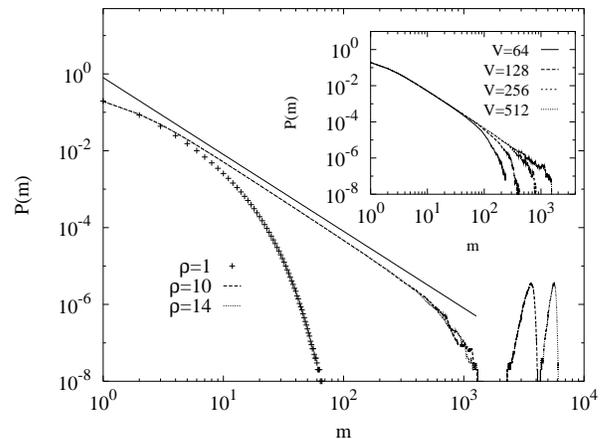}
 \caption{\label{fig1} $P(m)$ is plotted against $m$ for three values of
density. The cutoff is seen to depend on $\rho$ for small values of the
density while for larger values, it is independent of $\rho$. The mass in
excess of the critical mass forms an aggregate. The solid line has an
exponent $-2$. The simulation was done for system size $V=500$ and
$w=3.0$. In the inset, the dependence of the cut off of the power law on
$V$ is shown. The simulations were done for $\rho=15$.}
 \end{figure}

Bias is introduced in the motion of the masses on the lattice by letting
$p\neq q$. In the limits of only diffusion or only chipping, the steady
states reached are the same as those for the $p=q$ case. Hence, one might
expect a phase transition in the $\rho-w$ plane as before. Further, the
mean field analysis \cite{MKB1,MKB2} does not recognize any difference
between the two cases and thus predicts the same behavior as in the
zero-bias case.  Earlier Monte Carlo simulations of this system in one
dimension did indeed seem to suggest the existence of a phase transition
similar to the symmetric case, though with $\tau\approx 2.0$ \cite{MKB2}.
The results of these simulations are summarized in Fig.~\ref{fig1}. For a
fixed lattice size and $w$, when the total mass in the system is increased
beyond a certain critical mass, the formation of an aggregate is observed.
The power law regime has a lattice size dependent cutoff (see inset of
Fig.~\ref{fig1}) as in the symmetric case.  However, the value of $\tau$
($\sim 2.0$) is a puzzle since a finite system density implies that $\tau$
should be strictly greater than $2$ (the first moment, which is the
density of the system, would diverge if $\tau<2$).

In this paper, we analyze the model for $p\neq q$ and show that the
apparent existence of a phase transition in the Monte Carlo simulations is
purely a finite-size effect in one dimension, with the critical density
$\rho_c(V)$ for fixed $w$ diverging with system size $V$ as $\log(V)$. We
argue that the exponent $\tau$ is exactly $2$. In two and higher
dimensions however, a transition from an exponential to an aggregate phase
does occur at finite critical density.

The rest of the paper is organized as follows. In Sec.~\ref{sec2}, we show
analytically that an aggregate phase cannot exist in the one dimensional
system, thereby showing that there is no phase transition in one
dimension. Sec.~\ref{sec3} contains numerical evidence for the results of
Sec.~\ref{sec2}, from Monte Carlo simulations of the model.  In
Sec.~\ref{sec4} the exponents describing the probability distribution for
mass in one dimension are characterized.  Section~\ref{sec5} contains a
numerical study of the two dimensional problem.  Section~\ref{sec6}
contains a summary and conclusions. 

\section{\label{sec2} Arguments for no phase transition in the presence of
spatial bias in one dimension}

In this section, we prove that in the presence of a spatial bias, an
aggregate phase cannot be present at any finite density in one dimension. 
We do so by assuming that an aggregate phase exists, and then showing that
this leads to certain contradictions.  To proceed, assume that a single
infinite aggregate exists in the system and that the rest of the system is
at a finite critical density $\rho_c$ (analogous to the symmetric case). 
Consider now a frame of reference that is attached to this aggregate. Let
$\eta_{k}^{\pm}(t)$ be the mass transferred in an infinitesimal time $dt$
at time $t$ from a site $k$ lattice sites away from the aggregate to a
site $k\pm 1$ sites from the aggregate.  Then
 \be
 m_k (t+dt) = m_k(t) + \eta_{k-1}^{+} + \eta_{k+1}^{-} - \eta_{k}^{+} -
\eta_{k}^{-} + a_{k}^{+} + a_{k}^{-},
 \label{eq:1}
 \ee
 where $m_k(t)$ is the mass at a site $k$ lattice sites from the aggregate
at time $t$, and $a_k$ is the change in mass due to the diffusion of the
aggregate. The time dependence of the variables on the right hand side of
Eq.~(\ref{eq:1}) have been suppressed for the sake of clarity. Equation
~(\ref{eq:1}) accounts for all the ways in which the mass on a site $k$
sites away from the aggregate can change under the dynamics. These changes
are either caused by the mass transfer to and from neighboring sites as
exemplified by the $\eta_k$'s or by the motion of the aggregate, which
leads to a relabeling of sites, as depicted by the $a_k$'s.

From the definition of the model, it is clear that
 \be
 \eta_{k}^{+} = \cases{
 m_k & with prob $p dt$, \cr
 1-\delta_{m_k,0} & with prob $p w dt$, \cr
 0 & otherwise,\cr}
 \label{eq:2}
 \ee and similarly for $\eta_{k}^{-}$, with $p$ replaced by $q$ in
Eq.~(\ref{eq:2}). When the aggregate hops, the sites have to be relabeled
and this leads to
 \be
 a_{k}^{+} = \cases{
 m_{k+1} - m_{k} & with prob $p dt$,\cr
 0 & otherwise, \cr}
 \label{eq:3}
 \ee
 and correspondingly for $a_{k}^{-}$, with $k+1$ replaced by $k-1$ and $p$
by $q$ in Eq.~(\ref{eq:3}). Taking averages on both sides in
Eq.~(\ref{eq:1}) and setting the time derivatives to zero in the steady
state, we obtain
 \bea
 \lefteqn{\frac{d\rho_k}{dt}= 0 = w \left[ p s_{k-1}- (p+q) s_{k} +q
s_{k+1} \right] + (p+q) \times } \nonumber \\
 && \left[\rho_{k-1} - 2 \rho_{k} + \rho_{k+1} + \rho_0 (2 \delta_{k,0} -
\delta_{k,-1} - \delta_{k,1}) \right], \label{eq:4}
 \eea
 where $s_0=1$, and $\rho_k = \langle m_k \rangle$ and $s_k = \langle
1-\delta_{m_k,0} \rangle$ are the average density and occupation
probability of a site $k$ lattice sites away from the aggregate
respectively. A point to note about Eq.~(\ref{eq:4}) is that bias plays a
role only in the terms coupling to the chipping rate $w$.  It can be
checked that introducing a bias only in the diffusion move and keeping the
chipping move symmetric does not change the behavior of the model from the
fully symmetric version $p=q$.

We now consider Eq.~(\ref{eq:4}) in the steady state when the time
derivative is set to zero.  The set of linear equations in
Eq.~(\ref{eq:4}) may be solved for on a finite or an infinite lattice to
obtain the $\rho_k$'s in terms of the $s_k$'s. In the former case, a
closed form expression for the densities is easily obtained. However, it
is more informative to look at the equations for an infinite lattice,
where the sites with negative (left of aggregate) and positive (right of
aggregate) indices may be treated separately. For this case, we obtain for
$n>0$,
 \bea
 \rho_n &=& n \left[ \rho_1 + \frac{w}{p+q} (q s_1 -p) +
\frac{w(p-q)}{p+q} s_c^r \right] \nonumber \\
 &+& \frac{w (p-q)}{p+q} \sum_{k=1}^{n} (s_k - s_c^r) +\frac{w p}{p+q}
(1-s_n), \label{eq:5}\\
 \rho_{-n} &=& n \left[ \rho_{-1} + \frac{w}{p+q} (p s_{-1} -q) -
\frac{w(p-q)}{p+q} s_c^l \right] \nonumber \\
 &+& \frac{w (p-q)}{p+q} \sum_{k=-1}^{-n} (s_c^l - s_k)+ \frac{w q}{p+q}
(1-s_{-n}), \label{eq:6}
 \eea
 where $s_c^r= \lim_{n\rightarrow \infty} s_n$ and $s_c^l=
\lim_{n\rightarrow \infty} s_{-n}$. 

 By assumption, we require that $\rho_n$ tends to a finite value as
$n\rightarrow \infty$. For this, we require first that the term
proportional to $n$ in Eqs.~(\ref{eq:5}) and (\ref{eq:6}) vanishes {\it
and} secondly that $s_{\pm n}$ approaches its asymptotic value faster than
$1/n$. The first condition expresses $\rho_1$ and $\rho_{-1}$ in terms of
$s_1$ and $s_{-1}$ as
 \bea
 \rho_1 &=& \frac{ w (p-q s_1) - w (p-q) s_c^r}{p+q}, \label{eq:7}\\
 \rho_{-1} &=& \frac{ w (q-p s_{-1}) - w (q-p) s_c^l}{p+q} \label{eq:8},
 \eea
  which when substituted in the $k=0$ equation of Eq.~(\ref{eq:4}) leads
to
 \be
 s_c^r=s_c^l \equiv s_c. 
 \label{eq:9}
 \ee
 This is consistent with the fact that far from the aggregate, the
occupation probability is the same on either side.  The second condition,
that $s_{\pm n}$ approaches its asymptotic value faster than $1/n$, will
be useful for determining the exponents and will be studied numerically in
Sec.~\ref{sec3}. 

A important point to note from Eq.~(\ref{eq:5}) is that the two cases
$p=q$ and $ p \ne q$ are quite different.  For $p=q$, the cumulative sum
over the $s_n$'s vanishes and $\rho_n$ depends only on the site occupancy
of the site $n$.  However when $p \ne q$, the sum plays a role in
determining the value of the site density and it becomes important to
understand the behavior of $s_{\pm n}$ as a function of $n$.  We do that
in Sec.~\ref{sec3}. 

We now examine the two point correlations in the presence of the aggregate
in order to obtain further relations between the $\rho_k$'s and $s_k$'s.
Using these relations we will be able to show a contradiction.  Consider
the two point correlations in the aggregate frame of reference. Let
 \be
 C_{r,k} = \langle m_{r} m_{r+k} \rangle. 
 \label{eq:10}
 \ee
 The equations governing the temporal evolution of the two point mass-mass
correlations may be derived by considering the mass transfer between two
neighboring sites (see Eq.~(\ref{eq:1})). Multiplying together
Eq.~(\ref{eq:1}) for $m_{r}$ and the corresponding one for $m_{r+k}$,
keeping terms up to order $dt$, and taking averages, we obtain
 \begin{widetext}
 \bea
  \frac{d C_{r,k}}{dt} & =& p C_{r+1,k} + q C_{r+1,k-1} + p C_{r,k-1} + q
C_{r,k+1} + p C_{r-1,k+1} + q C_{r-1,k} - 3 (p+q) C_{r,k} \nonumber \\
 &&\mbox{} + w \left[ - p D_{r-1,k+1}+ (p+q) D_{r,k} - q D_{r+1,k-1}
 - p E_{r,k-1} + (p+q) E_{r,k} - q E_{r,k+1} \right] \nonumber \\ &&
\mbox{} - \delta_{k,1} \left[ q C_{r+1,0} + p C_{r,0} + w p s_r+ w q
s_{r+1} \right], \quad r,k =1,2,\ldots, \label{eq:11} \\
 \frac{d C_{r,0}}{dt} & =& (p+q) \left[ C_{r+1,0} - 2 C_{r,0} + C_{r-1,0}
\right] + 2 p C_{r-1,1} +2 q C_{r,1} - w \left[ 2 p D_{r-1,1} + 2 q
E_{r,1} \right] \nonumber \\
 && \mbox{} + w \left[ (p+q) s_r +p s_{r-1} + q s_{r+1}\right] \quad r =
 1,2,\ldots, \label{eq:12}
 \eea
 \end{widetext}
 where
 \bea
 D_{r,k}& = &\langle \delta_{m_{r} , 0} m_{r+k} \rangle \quad
k=1,2,\ldots, \label{eq:13}\\
 E_{r,k}& = &\langle m_r \delta_{m_{r+k} , 0} \rangle \quad k = 1,2,\ldots
.  \label{eq:14}
 \eea
 Also, for Eqs.~(\ref{eq:11}) and (\ref{eq:12}) to be valid when $r=1$, we
need to set $C_{0,k}=D_{0,k}\equiv 0$. In the steady state, the time
derivatives can be set to zero. Summing Eq.~(\ref{eq:11}) over all
$r,k=1,2,\dots$, and subtracting from this $1/2$ times the sum over $r =
1,2,\ldots$ in Eq.~(\ref{eq:12}), we obtain
 \be
 (p+q) C_{1,0} = w p - 2 (p+q) \sum_{k=1}^{\infty} C_{1,k} + 2 w q
\sum_{k=1}^{\infty} D_{1,k} + w q s_1. \label{eq:15}
 \ee
  The left hand side of Eq.~(\ref{eq:15}) is a finite number while the
right hand side has two infinite sums. These two sums can add up to give a
finite value only if asymptotically the terms in the two summation are
equal. Using the fact that the two point correlations decouple when the
separation between the two points become large, we obtain
 \be
 (p+q) \rho_1 = w q (1-s_1). 
 \label{eq:16}
 \ee
 Equation~(\ref{eq:16}) and (\ref{eq:7}) have to be simultaneously
satisfied. This is possible only if
 \be
 w (p-q) (1-s_c)=0. 
 \label{eq:17}
 \ee
 The fragmenting rate $w$ being equal to zero is the trivial limit in
which the entire mass in the system coagulates together to form an
infinite aggregate. When there is a bias $p\neq q$, then for finite $w$,
the only way Eq.~(\ref{eq:17}) can be satisfied is if $s_c=1$ ;  {\it
i.e}, the occupation probability far away from the aggregate is $1$. This
can occur only if $\rho_c=\infty$, which contradicts our initial
assumption that $\rho_c$ is finite. Another way of seeing a contradiction
is to consider Eq.~(\ref{eq:5}) when $n\rightarrow \infty$. Setting
$s_c=1$, we obtain that the densities far away from the aggregate become
negative for $p>q$. Thus, for any finite $w$ and $\rho$ we always have a
contradiction and hence our initial assumption of an aggregate existing at
finite density is proved wrong. This proves that for any finite $\rho$ and
$w$, an aggregate phase does not exist in one dimension in the
thermodynamic limit.  For the symmetric problem ($p=q$), there is no
contradiction between Eq.~(\ref{eq:16}) and (\ref{eq:7}), since
Eq.~(\ref{eq:17}) is automatically satisfied.

\section{\label{sec3} Numerical checks in one dimension}

The results of the previous section thus show that an infinite aggregate
cannot exist in an infinite system when there is a non zero bias. However,
numerical simulations of finite size systems do point to the existence of
an aggregate phase (see Fig.~\ref{fig1}). Nevertheless, there is no
contradiction with the results of Sec.~\ref{sec2}, provided the critical
density $\rho_c(V)$ diverges with the system size $V$. In this section, we
study numerically the system size dependence of $\rho_c(V)$ in one
dimension. 

\begin{figure}
 \includegraphics[width=8.0cm]{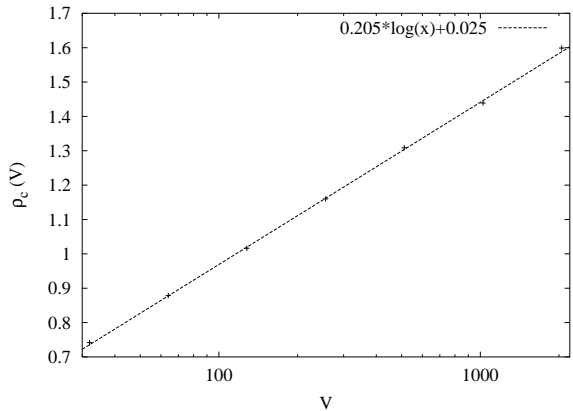}
 \caption{\label{fig2}$\rho_c(V)$ diverges with $V$ as $\log(V)$. The
simulations were done for the fully asymmetric model $p=1$ and $q=0$.}
 \end{figure}

In earlier studies\cite{MKB2}, the system size dependence of $\rho_c(V)$
was not investigated because there was no systematic way of making an
accurate numerical measurement of the critical density. To measure
$\rho_c(V)$, we adopted the following procedure. For a fixed lattice size,
we start the system with a density much higher than that required to form
an aggregate.  The system is then allowed to reach the steady state and
the biggest cluster is identified as the aggregate.  We then measure the
density in the rest of the system (excluding the aggregate) and use the
fact that the state of the rest of the system resembles that at
criticality. In Fig.~\ref{fig2}, the system size dependence of $\rho_c(V)$
is shown on a semi-log scale. From the numerical evidence, we conclude
that $\rho_c(V) \sim \log(V)$.

\begin{figure}
 \includegraphics[width=8.0cm]{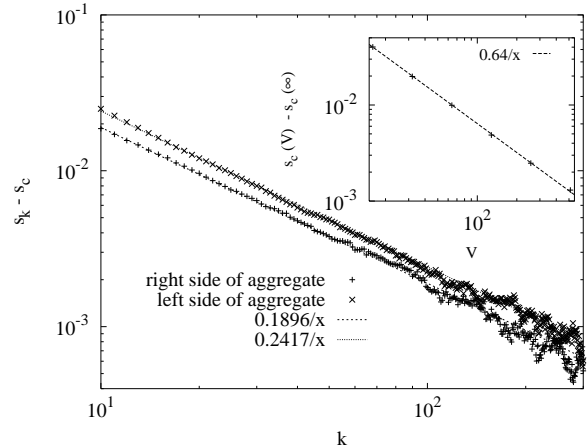}
 \caption{\label{fig3}$s_k$ converges to $s_c$ ($\approx 0.2775$ in this
case) as $1/k$. The simulations were done for lattice size 4000 and
$w=1.0$.  The inset shows the finite size correction to $s_c$.}
 \end{figure}

As a further check, we study the occupation probability numerically.  From
Eq.~(\ref{eq:5}), the dependence of $\rho_n$ on $\sum (s_{\pm k}- s_c)$,
taken together with the fact that $\rho_c(V) \sim \log(V)$ implies that
 \be
 |s_{\pm k}- s_c| \sim \frac{a_{\pm}}{|k|^x}, \quad k \gg 1,
 \label{eq:18}
 \ee
 with $x = 1$ which in turn implies than $s_c(V)$ converges to its
asymptotic value as $1/V$. Both of these requirements are consistent with
numerical simulations (see Fig.~\ref{fig3}). The simulations were done for
the fully asymmetric model $p=1$ and $q=0$. 

\section{\label{sec4} Probability distribution}

 \begin{figure}
 \includegraphics[width=8.0cm]{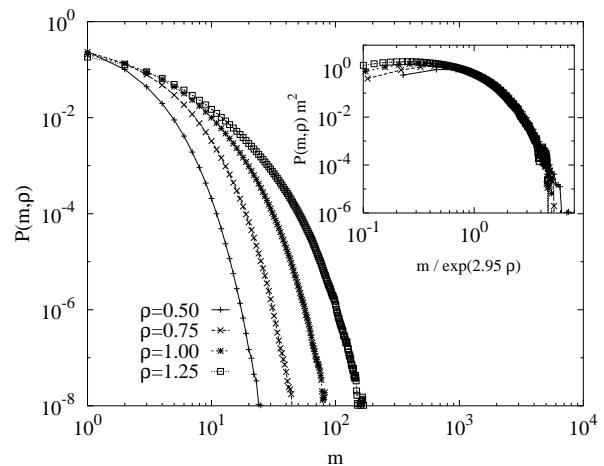}
 \caption{\label{fig4} The probability distribution for mass for different
values of $\rho$. The inset shows the scaling plots when the distributions
are scaled as in Eq.~(\ref{eq:20}).}
 \end{figure} From the analytical and numerical evidence of
sections~\ref{sec2} and \ref{sec3} it is clear that the system is
sensitive to the manner in which the limits $M \rightarrow \infty$ and $V
\rightarrow \infty$ are taken, where $M$ is the total mass in the system.
When $M$ is increased beyond $M_c$ ($=\rho_c V$) keeping $V$ fixed, an
infinite aggregate does form in the system.  In this regime, in analogy
with the symmetric problem \cite{RM}, we can then write a scaling form for
the probability distribution for $M>M_c$ as
 \be
 P(m,V) = \frac{1}{m^\tau} f\left(\frac{m} {V^\phi} \right) + \frac{1}{V}
\delta(m- (M-M_c)). 
 \label{eq:19}
 \ee
 Since the mean mass in the power law part of the distribution scales as
$\log(V)$ (see Fig.~\ref{fig2}), we immediately derive $\tau=2$. In
addition, from the consideration that there is only one aggregate, the two
exponents $\tau$ and $\phi$ are known to obey the scaling relation
$\phi(\tau -1)=1$ \cite{RM}. This implies that $\phi=1$.  The fact that
the cutoff of the power-law distribution scales as $V$, and not as a
smaller power of $V$ as in the symmetric case \cite{RM}, is consistent
with the fact that the transition does not exist for large $V$.

What is the behavior of the system when the order of limits is reversed
and $M, V \rightarrow \infty$ keeping $\rho = M/V$ fixed? In this case, we
make the reasonable ansatz,
 \be
 \lim_{V\rightarrow \infty} P(m,\rho) = \frac{1}{m^{\tau}} g\left(\frac{m}
{m^*} \right), \label{eq:20}
 \ee
 with the same exponent $\tau=2$. The requirement that $\langle m \rangle
= \rho$ taken together with Eq.~(\ref{eq:20}) implies that the cutoff
$m^{*} \sim e^{\alpha \rho}$.  Scaling plots of the probability
distribution for various values of $\rho$ scaled as in Eq.~(\ref{eq:20})
are shown in Fig.~\ref{fig4}.
 
From simulations, it is seen that the function $g(x) \sim \mbox{const}$
when $x\rightarrow 0$ (See inset of Fig.~\ref{fig4}). This taken together
with the exponential divergence of the mass cut-off implies that for
infinite $\rho$, $P(m)$ is a pure power-law.  This is similar to the
steady state of the Takayasu model \cite{takayasu} for river networks
where mass aggregates in the presence of a constant influx of particles,
and the steady state has a nontrivial power law distribution. 

\section{\label{sec5} Numerical Simulations in two dimensions}

\begin{figure}
 \includegraphics[width=8.0cm]{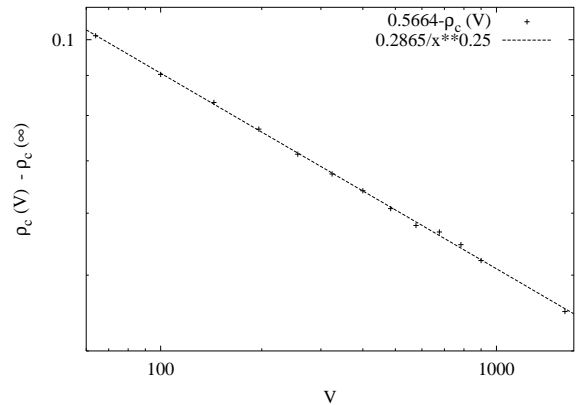}
 \caption{\label{fig5} The convergence of the critical density to a finite
value with increasing system size in two dimensions is shown.  The
simulations were done for the fully asymmetric model $p=1$, $q=0$ and the
fragmenting rate $w=1.0$.}
 \end{figure}

The arguments used in Sec.~\ref{sec2} to prove that there is no phase
transition in one dimension are very specific to one dimension and cannot
be extended to two and higher dimensions. Instead, in this section, we
numerically study the model in two dimensions.  First, we measure the
critical density $\rho_c(V)$ for lattice sizes varying from $8 \times 8$
to $32 \times 32$ using the same method as that used for the one
dimensional simulations.  In this case it is seen that $\rho_c(V)$ does
converge to a finite value when the system size is extrapolated to
infinity (see Fig.~\ref{fig5}). Hence, a phase transition does exist in
the infinite system limit. 
 \begin{figure}
 \includegraphics[width=8.0cm]{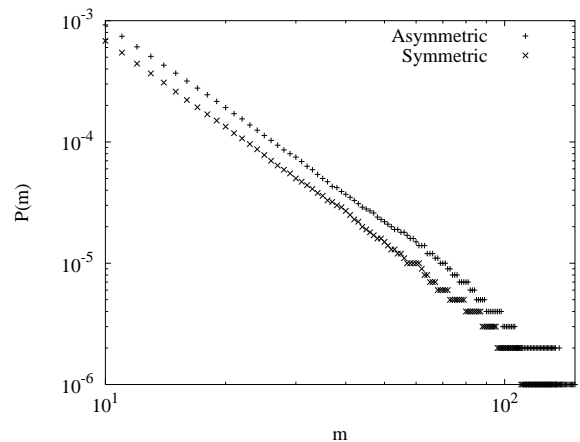}
 \caption{\label{fig6} The value of the power law exponent $\tau$ for the
fully asymmetric model in two dimensions is very close to that of the
symmetric model in two dimensions. The simulations were done for a
$30\times 30$ lattice with $w=1.0$ and $\rho=10.0$.}
 \end{figure}

We now show that for the system with spatial bias the value of the power
law exponent $\tau$ in two dimensions is close to the zero bias case. It
is difficult to make a direct measurement of $\tau$ because the cutoff to
the power law grows slowly with the system size and is not large enough
((cutoff $\sim 200$ when the system size $\sim 2000$)  for an accurate
measurement of the slope.  For instance in the zero bias case direct
measurements of $\tau$ \cite{MKB1,MKB2} gave a value close to $2.3$ while
indirect numerical methods \cite{RM} showed that $\tau$ is close to $5/2$
in all dimensions. Therefore, rather than measure $\tau$ directly for the
asymmetric model, we compare the simulation results (see Fig.~\ref{fig6})
for the fully asymmetric problem with those for the symmetric problem with
the same parameters. The slopes of the two curves are comparable. Hence,
we conclude that the exponent $\tau$ for the asymmetric model in two
dimensions is very close to $5/2$, as in the symmetric model. 

\section{\label{sec6} Summary and conclusions}

In summary, we have investigated in detail the effect of introducing a
spatial bias on the nonequilibrium phase transition in a model of
coagulation and fragmentation. We show analytically that the phase
transition is inhibited in one dimension. However a signature of the two
original phases remains and the scaling implications of this are
characterized. We have also resolved the puzzle of the exponent $\tau$
being very close to $2$. In two dimensions, the phase transition is shown
numerically to exist.

We now give a more intuitive explanation of why the phase transition gets
curbed in one dimension but not in two dimensions. In this model, there
are two competing processes.  While the diffusion move creates larger and
larger masses by coagulation, the fragmentation move tends to create
smaller masses as well as inhibit the formation of large masses. If the
diffusion move was to be considered by itself, then a cluster of size $l$
would be created in time of order $l^2$. In one dimension, if the
fragmentation move was to be considered by itself, then a fluctuation of
order $l$ would be dissipated in time of the order $l^{3/2}$ for the
asymmetric model and of the order $l^2$ for the symmetric model. These
exponents are known exactly because of the existing exact analogy in one
dimension. between only fragmentation with (without) bias and the
asymmetric (symmetric) exclusion process.  Thus, for the asymmetric
problem, fragmentation always wins out over diffusion and we only have an
exponential phase. However, in two dimensions, bias is irrelevant for the
fragmentation move and hence a fluctuation of order $l$ gets dissipated in
time of order $l^2$, which is of the same order as the time required to
create a cluster of size $l$ by diffusion. 

To carry this argument further, we can study the symmetric problem by
slowing down the diffusion rate. This can be done by introducing a mass
dependent diffusion rate $\sim m^{-\alpha}$ with $\alpha >0$. The above
arguments would then imply that this dynamics ought not to have a phase
transition for any $\alpha>0$.  This is indeed the case and it can be
shown that the phase transition does get curbed in all dimensions
\cite{RD}, as predicted. 

There remain several interesting questions to investigate further.  While,
we have numerically shown that $\tau = 2.0$, it would be interesting if it
could be derived from first principles by solving the model. Further, for
an infinite system, the probability distribution of the masses has the
form (Eq.~(\ref{eq:20}))
 \begin{eqnarray*}
 P(m) \sim \frac{1}{m^2} e^{-\beta m e^{-\alpha \rho}}, \quad m \gg 1,
 \end{eqnarray*}
 where $\beta$ could depend on $\rho$ and $w$ while $\alpha$ depends only
on $w$. The origin of the length scale $e^{\alpha \rho}$ under this
dynamics is an interesting point that remains to be understood. Also,
since the phase transition in one dimension is a finite size effect, the
implications of the traffic jam that was seen in \cite{IK} need to be
reexamined.

We would like to thank T.~Hanney, R.~Stinchcombe, G.~Sch\"utz and
O.~Zaboronski for useful discussions and S.~N.~Majumdar, M.~Barma and
D.~Dhar for a critical reading of an earlier version of the manuscript.
This work was supported by EPSRC, UK.

\end{document}